\documentclass[%
 reprint,
 amsmath,amssymb,
 aps,
]{revtex4-2}

\usepackage{graphicx}
\usepackage{dcolumn}
\usepackage{bm}
\usepackage[breaklinks=true,colorlinks=true,linkcolor=blue,urlcolor=blue,citecolor=blue]{hyperref}

\begin{document}

\preprint{APS/123-QED}

\title{Spin Squeezing through Collective Spin-Spin Interactions}

\author{Yanzhen Wang$^{1}$}
\thanks{These authors contributed equally to this work.}
\author{Xuanchen Zhang$^{1}$}
\thanks{These authors contributed equally to this work.}
\author{Yong-Chun Liu$^{1,2}$}%
 \email{ycliu@tsinghua.edu.cn}
\affiliation{%
 $^{1}$State Key Laboratory of Low-Dimensional Quantum Physics,\\
Department of Physics, Tsinghua University, Beijing 100084, P. R. China 
}%
\affiliation{
$^{2}$Frontier Science Center for Quantum Information, Beijing 100084, P.R. China
}%

%
%

\date{\today}

\begin{abstract}
Spin squeezing provides crucial quantum resource for quantum metrology and quantum information science. Here we propose that one axis-twisted (OAT) spin squeezing can be generated from free evolution under a general coupled-spin model with collective spin-spin interactions. We further propose pulse schemes to recover squeezing from parameter imperfections, and reach the extreme squeezing with Heisenberg-limited measurement precision scaling as $1/N$ for $N$ particles. This work provides a feasible method for generating extreme spin squeezing.
\end{abstract}

\maketitle


\section{Introduction}

The precision of quantum measurements with uncorrelated particles is restricted by the standard quantum limit, which can be overcome using nonclassical states \cite{Giovannetti2004,Pezze2018}. Among them, squeezed spin state \cite{Kitagawa1993,Wineland1994,Ma2011} is a promising candidate to improve precision by simply performing collective measurements. Various strategies have been proposed and demonstrated to generate spin squeezing, including quantum nondemolition measurements \cite{Kuzmich1998,Kuzmich2000,Appel2009,Anne2010,Bao2020} and dynamical evolution with nonlinear interactions. For the latter, an all-to-all two-body interaction, often referred to as one-axis twisting (OAT) model \cite{Kitagawa1993}, is widely studied to deterministically produce strong squeezing \cite{Jin2009,Pezze2009,Wu2015,Davis2016}. Even higher degree of squeezing is attainable with two-axis twisting (TAT) interaction \cite{Kitagawa1993}, which achieves Heisenberg-limited scaling of measurement precision \cite{Yukawa2014,Kajtoch2015}.

The squeezing interactions mentioned above require two-body coupling within the spin ensemble. Previous studies focus on several specific systems to create and witness such interactions. For example, the nonlinear atomic collisions in two-component Bose-Einstein condensate give rise to an OAT interaction \cite{Sorensen2001,Gross2010,Riedel2010}. In cavity QED systems, the OAT Hamiltonian can be engineered via dispersive atom-light interactions with cavity either driven \cite{Schleier2010,Leroux2010,Leroux2010_2,Zhang2015} or undriven \cite{Hu2017,Matthew2018,Lewis2018}, in which the atom-atom interaction is mediated by photons within the cavity. Another approach utilizes the strong interaction between Rydberg atoms to generate squeezing \cite{Gil2014} and is demonstrated in recent experiments \cite{Hines2023,Bornet2023,Eckner2023}. Other studies to produce OAT interaction exist in trapped ions \cite{Molmer1999,Britton2012,Justin2016} and lattice systems \cite{Sorensen1999,Kajtoch2018,He2019,Mamaev2021,Hernandez2022}. These existing researches are mainly restricted to certain systems and lack of universality. To further obtain the TAT interaction, potential strategies include implementing pulse sequences \cite{Liu2011,Shen2013,Zhang2014,Chen2019,Huang2021,Hu2023} and continuous driving field \cite{Huang2015} to the existing interactions, which still face great challenges of experimental realization.

In this work, we investigate a general model of collective interaction coupling two spin ensembles and show its ability of generating deterministic spin squeezing. When the number of particles in two ensembles differs greatly, the one with large particle number mediates an effective OAT interaction in the other ensemble. The imperfection arising from anisotropic coupling strength can be eliminated with spin echo sequences. We also propose a pulse scheme to synthesize the TAT interaction based on this model, resulting in the Heisenberg-limited squeezing. Compared with the previous proposal \cite{Huang2023}, the squeezing interaction here appears purely from the inter-species coupling, without requirement of additional driving fields. As a result, this scheme is easier for implementation.

\section{System Model}
Consider a system of two species of spin-1/2 (or two-level) particles, denoted by $S$ and $J$, with the inter-species coupling described by the following universal interaction Hamiltonian:
\begin{eqnarray}
H_{\text{int}} = g_x S_x J_x + g_y S_y J_y + g_z S_z J_z
\label{eq:1},
\end{eqnarray}
where $g_\mu$ $(\mu = x, y, z)$ denotes the coupling strength of different collective spin components between two species. $S_{\mu}=\frac{1}{2}\sum_{k=1}^{N_s}\sigma_{S,\mu}^{(k)}$ and $J_{\mu}=\frac{1}{2}\sum_{k=1}^{N_j}\sigma_{J,\mu}^{(k)}$ $(\mu = x,y,z)$ denotes the collective spin operators of the two subsystems, and can be viewed as angular momentum operators of total angular momenta $S = N_s/2$ and $J=N_j/2$, respectively. Here $\sigma_{S,\mu}^{(k)}$ and $\sigma_{J,\mu}^{(k)}$ denotes the Pauli matrices of the $k$-th particle within subsystem $S$ and $J$. $S_{\mu}$ and $J_{\mu}$ satisfy the commutation relations $[S_{i},S_{j}] = i\varepsilon_{ijk}S_k$ and $[J_{i},J_{j}] = i\varepsilon_{ijk}J_k$, where $\varepsilon_{ijk}$ $(i,j,k=x,y,z)$ is the Levi-Civita symbol. A wide range of spin-related coupling can be described by the Hamiltonian above, including spin-exchange interactions, $H_{\text{int}}=g\left(S_x J_x + S_y J_y + S_z J_z \right)$, and dipole-dipole interactions, $H_{\text{int}}=g\left(S_x J_x + S_y J_y - 2 S_z J_z \right)$. We choose the initial state to be the direct product of coherent spin state (CSS) in two subsystems, with the CSS in subsystem $J$ pointing at the $z$-axis, i.e., the eigenstate of $J_z$ with eigenvalue $N_j/2$ ($N_j$ is the corresponding particle number), and focus on the spin squeezing of subsystem (species) $S$, under the condition $J \gg S$.

We now show that such a system can generate OAT squeezing under free evolution. Since $J \gg S$, the quantum state of subsystem $J$ evolves much slower than the quantum state of subsystem $S$, meaning that in the time scale we are considering, the state has only very small deviations from $J_z=J$. One can therefore work in a subspace in the Hilbert space of our system, in which $H_{1}=g_x S_x J_x + g_y S_y J_y$ can be viewed as a perturbation to $H_0 = g_z S_z J_z$. To show the spin squeezing nature of such a system under free evolution, we perform a Frohlich-Nakajima transformation (FNT) \cite{Frohlich1950,Nakajima1955}: 
\begin{eqnarray}
H_{\text{FN}} &&= e^{-S_{\text{FN}}}H_{\text{int}}e^{S_{\text{FN}}} = H_{0} + H_1+[H_0,S_{\text{FN}}]\nonumber\\ 
&&+[H_1,S_{\text{FN}}]+\frac{1}{2}[[H_0,S_{\text{FN}}],S_{\text{FN}}]+...
\label{eq:FN}
\end{eqnarray}
where 
\begin{eqnarray}
S_{FN}&&=\frac{1}{4g_z J}\big[(g_x-g_y)\left(S_-J_- -S_+J_+\right)\nonumber\\
&&+(g_x+g_y)\left(S_-J_+ - S_+J_- \right)\big]
\label{eq:S},
\end{eqnarray}
and $J_{\pm}=J_x\pm iJ_y$. We then apply the Holstein-Primakoff transformation \cite{Holstein1940}:
\begin{eqnarray}
J_{+} &&= \sqrt{2J-a^{\dagger}a}a,\ J_{-}=a^{\dagger}\sqrt{2J-a^{\dagger}a},\nonumber\\
J_z &&= J-a^{\dagger} a
\label{eq:2}.
\end{eqnarray}
Here, $a^{\dagger}$ and $a$ are bosonic creation and annihilation operators satisfying $[a,a^{\dagger}]=1$. 
In the subspace under consideration, $J_z$ is close to $J$, i.e., $\langle a^{\dagger}a\rangle \ll J$, so we can make the approximation $J_+\approx \sqrt{2J}a,\ J_-\approx \sqrt{2J}a^{\dagger}$. By using of this approximation, we obtain the effective Hamiltonian for the coupled system:
\begin{widetext}
\begin{eqnarray}
H_{\text{FN}} &&\approx \left[g_z J + \frac{g_x^2+g_y^2}{4g_z} +\frac{g_x^2 - g_y^2}{4g_z}(aa+a^{\dagger}a^{\dagger}) + \frac{g_x^2+g_y^2-2g_z^2}{2g_z}a^{\dagger}a \right]S_z \nonumber\\
&&+\frac{g_xS_x(a+a^{\dagger})+ig_y S_y(a^{\dagger}-a)}{\sqrt{2J}}+\frac{g_xg_y}{2g_z} S_z^2  + ...
\label{eq:FNapprox}
\end{eqnarray}
\end{widetext}
Neglect high order terms in $N_s/N_j$ and $\langle a^{\dagger}a \rangle / J$, the effective Hamiltonian is reduced to
\begin{equation}
H_{\text{FN}} \approx g_z J S_z + \frac{g_x g_y}{2g_z}S_z^2
\label{eq:eff},
\end{equation}
which is an OAT Hamiltonian with a linear term.

This effective Hamiltonian can be understood in two ways. Firstly, from the mathematical point of view, the system is evolving in the subspace in which $\langle a^{\dagger} a \rangle \ll J$, so the matrix elements of $S_{\text{FN}}$ is much smaller than that of $H_{\text{int}}$ in this subspace. Therefore, (6) means that, in this subspace, $H_{\text{FN}}$ and the original interaction Hamiltonian $H_{\text{int}}$ are very close to each other. Secondly, one can view the subsystem $J$ as an intermediary. When spin $S$ evolves and perturbs the large spin $J$, such perturbation induces a ``back action" on spin $S$, mediating its inter-species interaction. This mechanism is similar to that in \cite{Huang2023}, and is analogous to the excitation in condensed-matter spin waves \cite{Holstein1940}. However, different from the previous work, no external control is needed to produce such a "back action" mechanism, making our proposal a more feasible squeezing scheme.\\

\section{\label{sec:numerical}Numerical Investigation}
\subsection{\label{subsection:3.1}OAT squeezing dynamics}

\begin{figure}
\includegraphics[width=0.5\textwidth]{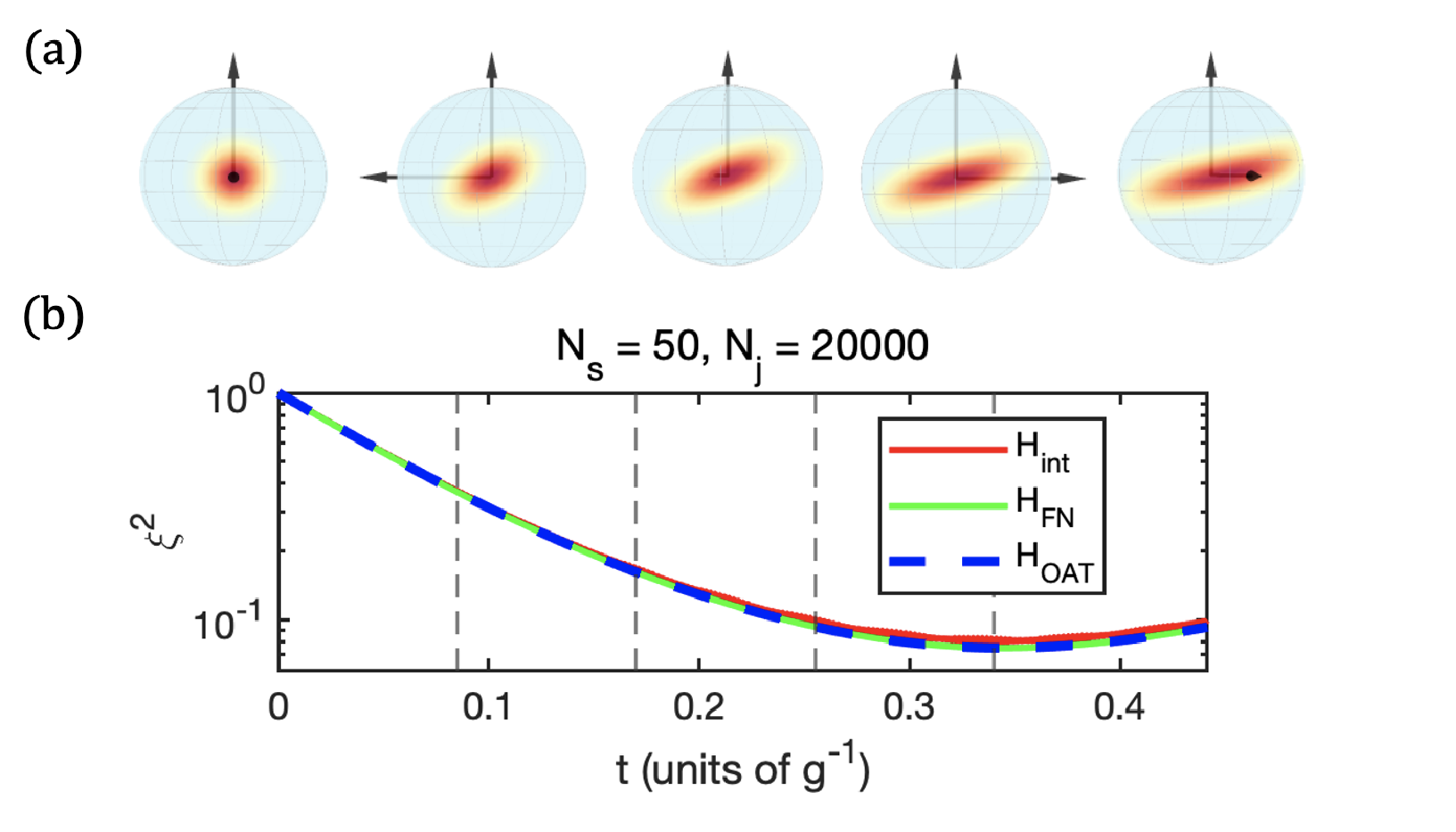}
\caption{\label{fig:fig1} Evolution of the quantum state under squeezing Hamiltonian. (a) Evolution of the quantum
state for spin $S$ at the time instants denoted by the vertical
gray dashed lines in (b), represented by the Husimi Q function on the generalized Bloch spheres. The horizontal and vertical arrows represent the $x$ and $z$ axes, respectively. (b) The free evolution of the squeezing parameters $\xi^2$ for subsystem $S$, for $H_{\text{int}}$ (red solid curve), $H_{\text{FN}}$ (green solid curve) and $H_{\text{OAT}}$ (blue dashed curve) defined in the main text, with $g_x = g_y = g$, $g_z = -2g$, $g>0$. The vertical gray dashed lines indicate time instants of $t = t_{\text{min}}/4$, $t_{\text{min}}/2$, $3t_{\text{min}}/4$ and $t_{\text{min}}$, with $t_{\text{min}}$ being the optimal squeezing time of OAT. The particle numbers are $N_s = 50$ and $N_j = 20000$.}
\end{figure}

\begin{figure}
\includegraphics[width=0.5\textwidth]{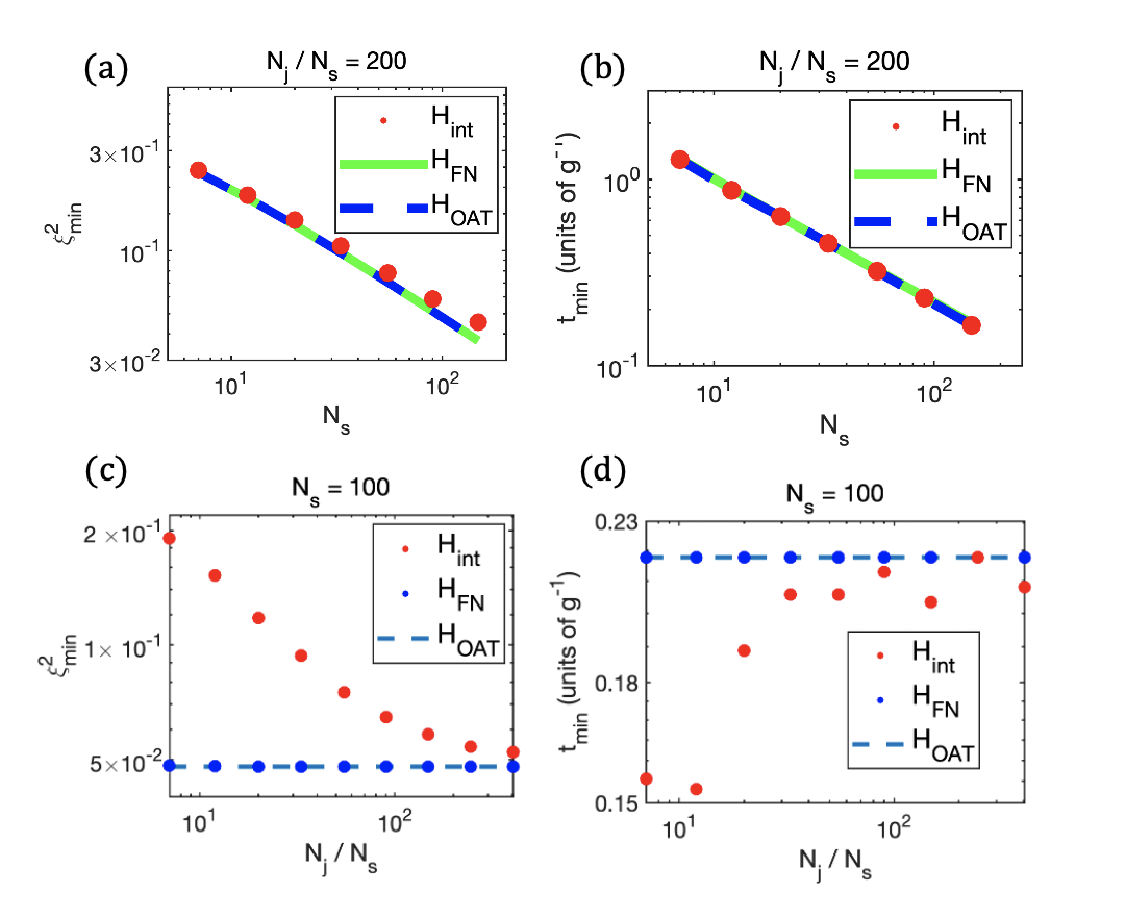}
\caption{\label{fig:fig2} Scaling analysis of the effective OAT squeezing under Hamiltonian $H_{\text{int}}$, with $g_x = g_y = g$, $g_z = -2g$. (a) and (b): The optimal spin squeezing $\xi^2_{\text{min}}$ and the optimal squeezing time $t_{\text{min}}$ as functions of the particle number $N_s$ for free evolution under $H_{\text{int}}$ (red solid balls), compared with the transformed Hamiltonian $H_{\text{FN}}$ (green solid lines) and the effective OAT results (blue dashed lines). (c) and (d): The optimal spin squeezing $\xi^2_{\text{min}}$ and the optimal squeezing time $t_{\text{min}}$ as functions of $N_j/N_s$ for free evolution under $H_{\text{int}}$ (red solid balls), compared with the transformed Hamiltonian $H_{\text{FN}}$ (blue solid balls) and the effective OAT results (blue dashed lines). The particle numbers are given in each subgraph.}
\end{figure}

To verify the validity of OAT squeezing under free evolution of the Hamiltonian Eq.~(\ref{eq:1}), we numerically investigate the evolution of the quantum state. As is illustrated by the Husimi Q representation on the generalized Bloch Spheres(after tracing out subsystem $J$), the quasiprobability distribution of subsystem $S$ is continuously squeezed, signifying the occurrence of spin squeezing, as depicted in Fig.~\ref{fig:fig1}(a). The degree of spin squeezing is usually characterized by the squeezing parameter $\xi^2=4\left(\Delta S_{\perp}\right)_{\min }^2 / N_s$, where
$\left(\Delta S_{\perp}\right)_{\min }^2$ is the minimum of the fluctuation $\left(\Delta S_{\perp}\right)^2=\left\langle S_{\perp}^2\right\rangle-\left\langle S_{\perp}\right\rangle^2$ for the spin component perpendicular to the mean spin direction. In Fig.~\ref{fig:fig1}(b) we compare the squeezing parameters for free evolution under $H_{\text{int}}$, $H_{\text{FN}}$ and the OAT Hamiltonian 
\begin{eqnarray}
H_{\text{OAT}}=\frac{g_x g_y}{2 g_z} S_z^2
\label{eq:OAT}.
\end{eqnarray}
The agreement of the evolution of squeezing parameters illustrates the validity of the effective OAT squeezing. Such validity is further demonstrated in Fig.~\ref{fig:fig2}(a) and (b), where the power-law scaling of optimal spin squeezing parameter $\xi^2_{\text{min}}$ and the corresponding squeezing time $t_{\text{min}}$ with respect to system size are compared with ideal OAT squeezing, and we find a clear consistency. For ideal OAT~(\ref{eq:OAT}), the scaling relation is given by \cite{Jin2009}
\begin{eqnarray}
\xi_{\min }^2 \simeq \frac{1}{2}\left(\frac{N_s}{3}\right)^{-\frac{2}{3}},\ t_{\min } \simeq \frac{2 \times 3^{1 / 6} |g_z|}{|g_x g_y| N_s^{2 / 3}}
\label{eq:OATscaling}.
\end{eqnarray}

We have also studied how the size of subsystem $J$ affects the quality of squeezing. Fig.~\ref{fig:fig2}(c) and (d) shows the relation between $N_j/N_s$, the ratio between subsystem sizes, and $\xi^2_{\text{min}}$ and $t_{\text{min}}$. We observe that the two quantities approaches the ideal OAT value when $N_j/N_s$ increases, as expected from the condition $J\gg S$, and a ratio of the order $10^2$ is sufficient for our model to produce squeezing that closely resembles ideal OAT dynamics.

\begin{figure}
\includegraphics[width=0.45\textwidth]{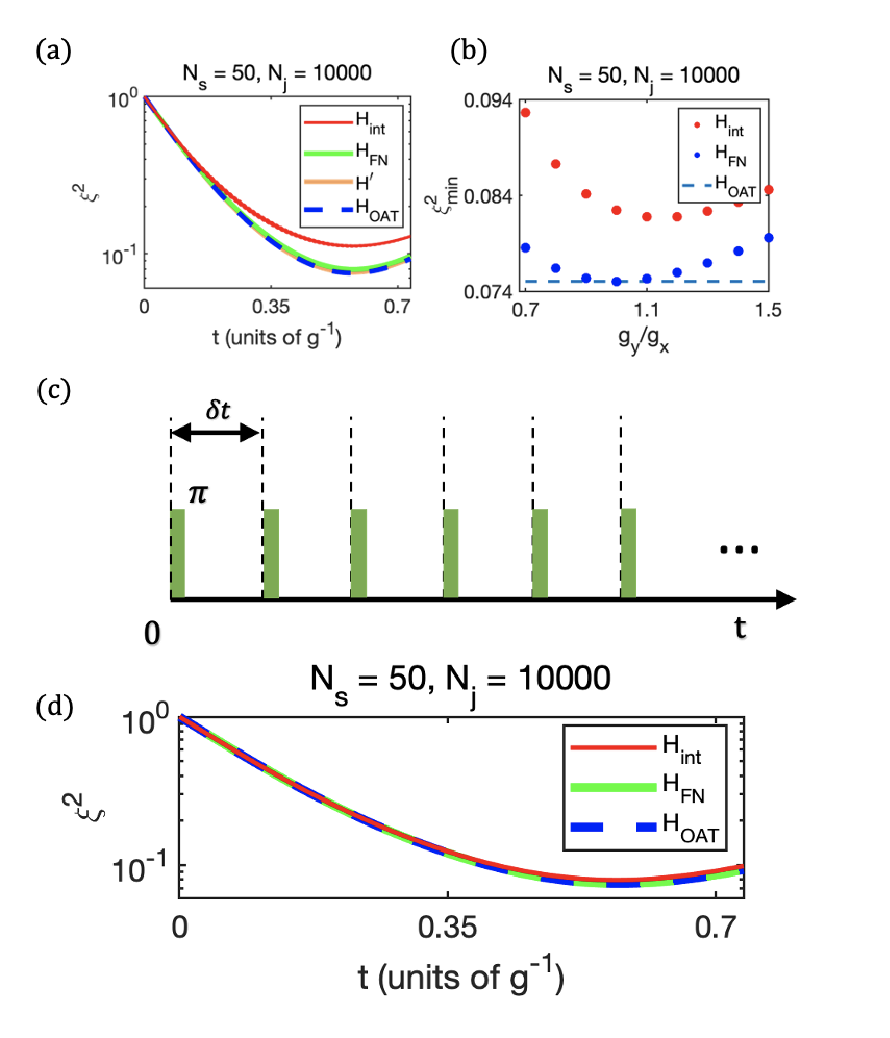}
\caption{\label{fig:fig3} OAT squeezing for $g_x \neq g_y$. (a) Evolution of the squeezing parameter $\xi^2$ under Hamiltonians $H_{\text{int}}$ (red solid curve), $H_{\text{FN}}$ (green solid curve), $H^{\prime}$ (golden solid curve) and $H_{\text{OAT}}$ (blue dashed curve). (b) The optimal squeezing parameter $\xi^2_{\text{min}}$ as a function of $g_y/g_x$ for $H_{\text{int}}$ (red solid dots) and $H_{\text{FN}}$ (blue solid dots), compared with the optimal squeezing parameter for ideal OAT (blue dashed line). (c) Pulse scheme for recovering OAT squeezing, consisting of $\pi$ rotation pulses (green rectangles) acting on subsystem $S$ equally spaced in time. (d) Evolution of the squeezing parameters $\xi^2$ for the OAT pulse scheme (red solid curve) and for the transformed Hamiltonian Eq.~(\ref{eq:FN}) (green solid curve), compared with the effective OAT dynamics Eq.~(\ref{eq:OAT}) (blue dashed curve). The model parameters are chosen such that $g_x^2+g_y^2=2g^2$ and $g_z = -2g$ are fixed. In (a) and (c), $g_y / g_x = 0.6$, $g_z = -2g$, and the particle numbers are $N_s = 50$ and $N_j = 10000$. The pulse separation is chosen to be $\delta t = 4\pi/\left(N_j |g_z| \right)$ to minimize simulation error, and the number of pulses until optimal squeezing is 900.}
\end{figure}

When $g_x \neq g_y$, numerical simulations indicate that, given the same $N_j/N_s$ ratio as in $g_x=g_y$ case, while $H_{\text{FN}}$ still yields OAT-level squeezing, the squeezing ratio of the quantum state evolving under the original Hamiltonian $H_{\text{int}}$ deviates from OAT, as shown in  Fig.~\ref{fig:fig3}(a) and (b). The following observation, however, shows that this imperfection can be overcome by making use of the famous spin echo pulse sequence \cite{Hahn1950}. Fig.~\ref{fig:fig3}(a) compared the the squeezing parameters for free evolution under $H_{\text{int}}= e^{S_{\text{FN}}}H_{\text{FN}}e^{-S_{\text{FN}}}$, $H^{\prime} = e^{S_{\text{FN}}}H_{\text{OAT}}e^{-S_{\text{FN}}}$ and $H_{\text{OAT}}$. The discrepancy between the first two indicates that the deviation of $H_{\text{int}}$ from OAT squeezing is mainly caused by the difference between $H_{\text{FN}}$ and $H_{\text{OAT}}$, which is approximately the linear-in-$S_z$ term in the RHS of Eq. (\ref{eq:FNapprox}) up to lowest order. Therefore, the imperfection can be eliminated by cancelling out the effect of this term. This can be attained by applying a sequence of $\pi$ pulses to system $S$ equally spaced in time, as illustrated in Fig.~\ref{fig:fig3}(c). The numerical simulation of squeezing dynamics shown in Fig.~\ref{fig:fig3}(d) verifies the validity of the pulse scheme.

\subsection{Generating TAT squeezing from effective OAT}
\begin{figure}
\includegraphics[width=0.47\textwidth]{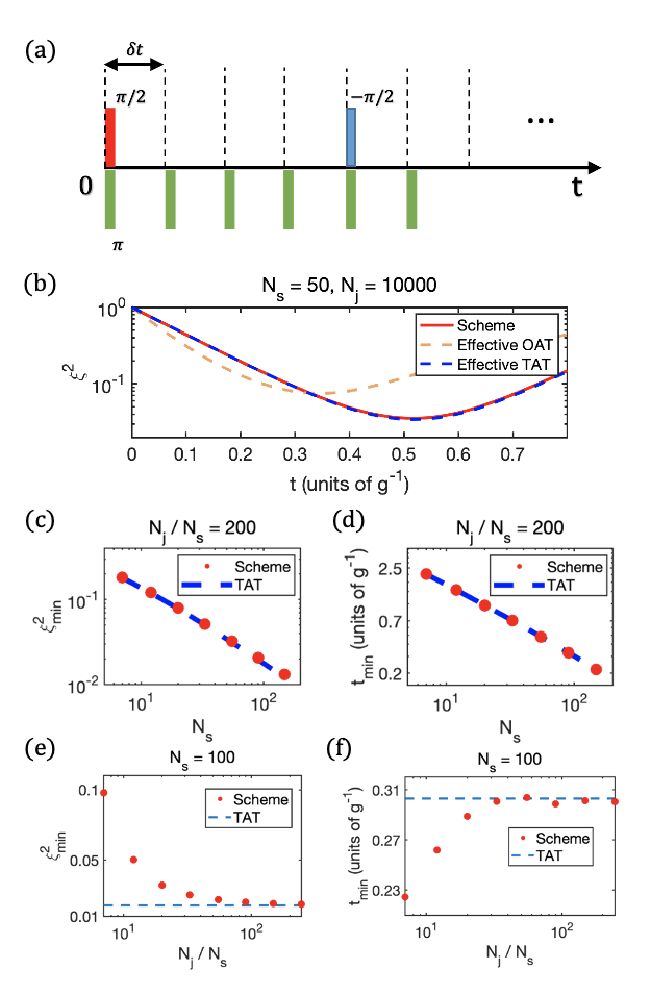}
\caption{\label{fig:fig4} Synthesis of TAT squeezing under $H_{\text{int}}=g\left(S_x J_x + S_y J_y - 2 S_z J_z \right)$. (a) An illustration of the pulse sequence. Each period $T_c = 6\delta t$ contains six $\pi$ rotation pulses (green rectangles), one $\pi/2$ pulse (red rectangle) and one $-\pi/2$ pulse (blue rectangle) along the $y$ axis imposed on subsystem $S$. (b) Evolution of the squeezing parameters $\xi^2$ for the TAT pulse scheme (red solid curve) and for the effective TAT Hamiltonian Eq.~(\ref{eq:TAT}) (blue dashed curve), compared with the OAT dynamics Eq.~(\ref{eq:OAT}) (golden dashed curve). (c) and (d): The optimal spin squeezing $\xi^2_{\text{min}}$ and the optimal squeezing time $t_{\text{min}}$ as functions of the particle number $N_s$ for the TAT pulse scheme (red solid balls), compared with the effective TAT results (blue dashed curves). (e) and (f): The optimal spin squeezing $\xi^2_{\text{min}}$ and the optimal squeezing time $t_{\text{min}}$ as functions of $N_j/N_s$ for the TAT pulse scheme (red solid balls), compared with the effective TAT results (blue dashed lines). The particle numbers are given in each subgraph. The pulse separation is chosen to be $\delta t = 4\pi/\left(N_j |g_z| \right)$ to minimize simulation error, and the number of pulses until optimal squeezing is 822.}
\end{figure}

Inspired by the transformation scheme from OAT to TAT in ref.~\cite{Liu2011}, starting from the Hamiltonian $H_{\text{int}}$ that is effectively OAT, squeezing can be further enhanced by applying a more intricate pulse sequence to generate TAT interaction. This scheme is demonstrated in Fig.~\ref{fig:fig4}(a). It is essentially a combination of the pulse sequence described in Fig.~\ref{fig:fig3}(c), which eliminates the linear-in-$S_z$ term in $H_{\text{eff}}$, and the pulse sequence in ref.~\cite{Liu2011}, which synthesizes a TAT interaction from the OAT Hamiltonian $H_{\text{OAT}}$. Define the rotation operator $R_{\theta} = e^{-i\theta S_y}$ and the evolution operator $U(\delta t) = e^{-iH_{\text{int}}\delta t}$, The evolution operator for a single period $T_c=6\delta t$ is given by
\begin{eqnarray}
U_1(6\delta t) = \left[  U\left(\delta t \right) R_{\pi}\right]^2 R_{-\pi/2} \left[  U\left(\delta t \right) R_{\pi}\right]^4 R_{\pi/2}
\label{eq:tatscheme}.
\end{eqnarray}
For $N_j \gg N_s$, $H_{\text{int}}\approx H_{\text{eff}}$. Using the identity
\begin{eqnarray}
\left[ e^{-i\left(AS_z+BS_z^2 \right)}R_{\pi} \right]^2 = e^{-2i B S_z^2}
\label{eq:pulseid1}
\end{eqnarray}
and the result in ref.~\cite{Liu2011}:
\begin{eqnarray}
e^{-i\tau S_z^2} R_{-\pi/2} e^{-2i\tau S_z^2} R_{\pi/2} \approx e^{-i\tau \left(2J_x^2 + J_z^2 \right)}
\label{eq:pulseid2}
\end{eqnarray}
for $\tau \ll \left(2N_s\right)^{-1}$, we arrive at
\begin{equation}
U_1(6\delta t) \approx e^{-i g_x g_y \delta t \left( 2S_x^2+S_z^2 \right)/g_z}
\label{eq:pulseid3}
\end{equation}
for $\delta t \ll \big|{g_z}/\left({g_x g_y N_s}\right)\big|$, which is equivalent to the evolution operator given by
\begin{equation}
H_{\text{TAT}} = \frac{g_x g_y}{6g_z}(2S_x^2 + S_z^2) = \frac{g_x g_y}{6g_z}(S_x^2 - S_y^2)
\label{eq:TAT}
\end{equation}
(the second identity is obtained by subtracting a constant $S(S+1)$). 

The validity of our TAT scheme is demonstrated by the numerical simulation in Fig.~\ref{fig:fig4}(b)-(f). Fig.~\ref{fig:fig4}(b) illustrates the evolution of the spin squeezing parameter under the pulse scheme, reflecting well alignment with that of the effective TAT Hamiltonian and a substantial outperformance compared with OAT squeezing. The optimal squeezing parameter and the corresponding squeezing time for the effective TAT interaction are approximately given by
\begin{equation}
\xi_{\min }^2 \simeq \frac{1.8}{N_s},\ t_{\min} \simeq \frac{3 |g_z|\ln (4N_s)}{|g_x g_y| N_s}
\label{eq:TATscaling},
\end{equation}
as is verified in Fig.~\ref{fig:fig4}(c) and (d). We have also investigated the trend of $\xi^2_{\text{min}}$ and $t_{\text{min}}$ as $N_j/N_s$ increases when $N_s$ is fixed. The result is shown in Fig.~\ref{fig:fig4}(e), (f), indicating that the two parameters agrees well with TAT for $N_j/N_s > 100$.

\begin{figure}[t]
\includegraphics[width=0.5\textwidth]{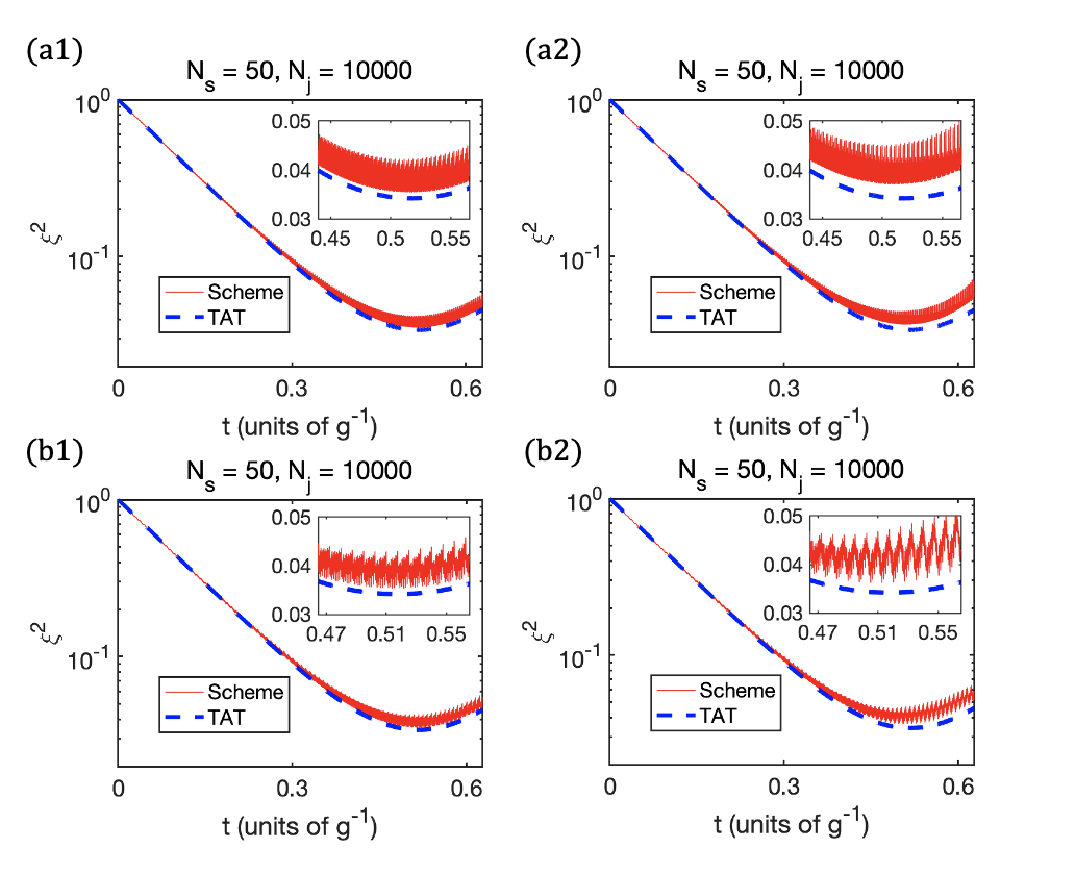}
\caption{\label{fig:fig5} Numerical noise analysis of the TAT pulse scheme with $N_s = 50, N_j=10000$. The first row show the evolution of the squeezing parameter under (a1) 0.01\% and (a2) 0.05\% of Gaussian stochastic noise adding on the pulse area. The second row show the squeezing parameter under (b1) 0.01\% and (b2) 0.05\% of Gaussian stochastic noise adding on the pulse separation. Inset: zoom in of the evolution curves near optimal squeezing.} 
\end{figure}

Our pulse scheme for generating TAT spin squeezing is subjected to various noises, including imperfections in pulse areas and pulse separations. To investigate the effect of such fluctuations, we simulate the squeezing dynamics of the pulse scheme by adding Gaussian stochastic noises, i.e., assuming the pulse areas or pulse separations are independent and subject to Gaussian distribution. As is shown in Fig.~\ref{fig:fig5}, under noise in the pulse area and pulse separation less than 0.05\%, our scheme can attain almost optimal squeezing of the effective TAT dynamics, indicating the robustness of our scheme.

\section{Conclusion}
In summary, we have proposed that spin squeezing can be generated from systems with collective spin-spin interaction. When the size of two spin subsystems differ greatly ($N_j/N_s \gtrsim 10^2$), effective OAT squeezing can be generated under free evolution, without any additional control. The influence of asymmetric coupling strengths on the squeezing can be removed by using a spin-echo pulse sequence consisting of equally-spaced $\pi$ pulses. Effective TAT squeezing with Heisenberg-limited measurement precision can be realized by applying a more intricate periodic pulse scheme, which incorporates spin echo and OAT-to-TAT transform. The pulse scheme is robust against pulse area and pulse separation fluctuations. This work provides a viable approach for generating high level of spin squeezing in a broad range of systems.

\begin{acknowledgments}
Y.W. thanks Jinyu Liu and Qixian Li for enlightening discussions. This work is supported by the National Key R\&D Program of China (Grant No. 2023YFA1407600), and the National Natural Science Foundation of China (NSFC) (Grants No. 12275145, No. 92050110, No. 91736106, No. 11674390, and No. 91836302).
\end{acknowledgments}



\nocite{*}

\bibliography{SJsqueezing}

\end{document}